\documentclass[preprintnumbers,pre]{revtex4}

\usepackage{amssymb}
\usepackage{amsmath}
\usepackage{graphicx}

\begin{document}

\title{Solution of the Percus-Yevick equation for hard discs}

\author{M. Adda-Bedia\dag, E. Katzav\dag, and D. Vella\dag\ddag}

\affiliation{\dag Laboratoire de Physique Statistique de l'Ecole Normale Sup\'erieure, CNRS UMR8550,
24 rue Lhomond, 75231 Paris Cedex 05, France.\\
\ddag Department of Applied Mathematics and Theoretical Physics, University of Cambridge, Wilberforce Road, Cambridge CB3 0WA, United Kingdom.}

\date{\today}

\begin{abstract}

We solve the Percus-Yevick equation in two dimensions by reducing it to a set of simple integral equations. We numerically obtain both the pair correlation function and the equation of state for a hard disc fluid and find good agreement with available Monte-Carlo calculations. The present method of resolution may be generalized to any even dimension.

\end{abstract}

\maketitle

\section{Introduction}

The description of hard-core fluids by approximate theories has attracted a lot of attention throughout the years \cite{hansen}. The reason for this interest is twofold: firstly a solution of the full problem was, and still is, extremely difficult. Secondly, approximate methods have allowed for very good predictions in the low density phase. Among the most widely used approximations is the Percus-Yevick (PY) equation for $d$-dimensional  hard spheres \cite{PY}.

Baxter developed a powerful method for solving the PY equation that works in odd dimensions \cite{baxter}. Leutheusser used this method to reduce the nonlinear integral PY equation to a set of nonlinear algebraic equations of order $d - 3$ (for $d>3$) \cite{leuth1}. However, this set of equations can only be solved analytically for odd $d\leq7$. In general, even the numerical solution of Leutheusser's equations is difficult to obtain  for higher dimensions because a general solution to nonlinear algebraic equations of order larger than four is not available. Analytical results have been found for one \cite{wertheim1}, three \cite{wertheim2,thiele}, five \cite{leuth1} and, most recently, seven dimensions \cite{robles}.

In two dimensions an approximate numerical solution of the PY equation was found by Lado~\cite{lado}. Leutheusser \cite{leuth2} was able to fit many of Lado's results using an ansatz for the direct correlation function. Rosenfeld \cite{rosenfeld} generalized Leutheusser's Ansatz to higher dimensions and compared the results with the analytical results in three and five dimensions. All other results available in the literature are based on Molecular Dynamics (MD) or Monte Carlo (MC) methods \cite{hoover67,numerics,alder68,wood,numerics2,clisby06}.

In this paper, we solve the PY equation for hard discs ($d=2$). We develop a method that reduces the problem to a set of integral equations that are solved numerically without major difficulties. The originality of the present method is based on techniques borrowed from the resolution of crack problems \cite{sned} and uses some results from Baxter's classical method~\cite{hansen,baxter}. The main difference from the hard sphere case (and any odd dimension in general) is that the problem of finding the indirect correlation function and the direct correlation function are coupled. This means that the present analysis necessarily yields both correlation functions and therefore provides the equation of state. Note that Lado~\cite{lado} solved the problem using approximate integral equations, in contrast to our exact integral equations. The advantage of the current method is that one can improve the precision at will, and that it may be generalized to higher even dimensions.

\section{The PY approximation}

The pair correlation function $g({\mathbf r})$ is related to the direct correlation function $c({\mathbf r})$ through the Ornstein-Zernike equation by \cite{hansen,OZ}
\begin{equation}
\label{eq:O&Z}
h({\mathbf r})=c({\mathbf r})+\rho \int_{0}^{\infty} h({\mathbf r'})c(|\mathbf{ r}-\mathbf{ r'}|) dr'\,,
\end{equation}
where $\rho$ is the particle number density and
\begin{equation}
h({\mathbf r})=g({\mathbf r})-1\,,
\end{equation}
is the indirect correlation function. The PY approximation is a closure relation of Eq.~(\ref{eq:O&Z}). For a hard-core pair interaction potential, this approximation reads \cite{hansen}
\begin{eqnarray}
g(r) & = & h(r)+1=0, \qquad r<1 \;,
\label{eq:bc1}\\
c(r) & = & 0,\qquad r>1 \;.
\label{eq:bc2}
\end{eqnarray}
Here and elsewhere, we use the radius of the sphere as the unit of length. Thus in two dimensions, we have $\rho=4\eta/\pi$ where $\eta$ is the packing fraction, and the space filling density corresponds to $\eta=1$. We define the two-dimensional Fourier transform
\begin{equation}
\tilde{F}(q)=2\pi\int_{0}^{\infty}rJ_0(qr)F(r)dr\,,
\label{eq:fourier}
\end{equation}
where $J_0$ is the zeroth order Bessel function. The inverse Fourier transform is then given by
\begin{equation}
F(r)=(2\pi)^{-1}\int_{0}^{\infty}qJ_0(qr)\tilde{F}(q)dq\,.
\label{eq:fourier1}
\end{equation}
In Fourier space Eq.~(\ref{eq:O&Z}) reads
\begin{equation}
\label{eq:O&Z1}
\left[1-\rho \tilde{c}(q)\right]\left[1+\rho \tilde{h}(q)\right]=1\,,
\end{equation}
Finally, the static structure factor $s(q)$ of wavenumber ${\mathbf q}$ is related to the pair correlation function through
\begin{equation}
s(q)=1+\rho \tilde{h}(q)\equiv\frac{1}{1-\rho \tilde{c}(q)}\,.
\end{equation}

\section{Resolution}

Condition (\ref {eq:bc2}) together with Eq.~(\ref {eq:fourier1}) imposes that $\tilde{c}(q)$ can be written as \cite{sned}
\begin{equation}
\tilde{c}(q)=\frac{1}{q}\int_0^1 \phi(t)\sin qt dt\,,
\label{eq:solcq}
\end{equation}
where $\phi(t)$ is a real function. Thus using Eq.~(\ref {eq:fourier1}), $c(r)$ is expressed as \cite{grads}
\begin{equation}
c(r)= \int_r^1 \frac{\phi(t)}{\sqrt{t^2-r^2}} \frac{dt}{2\pi}\,.
\label{eq:solc}
\end{equation}
The function $c(r)$ does not diverge at $r=1^-$, thus the function $\phi(t)$ satisfies
\begin{equation}
\phi(t)\simeq \frac{4c(1^-) t}{\sqrt{1-t^2}} \qquad \mbox{as}\;\;t\rightarrow 1^-\,.
\label{eq:asympt}
\end{equation}
Define
\begin{equation}
\frac{1}{s(q)}=A(q)=1-\rho \tilde{c}(q)= 1-\frac{\rho}{q} \int_0^1 \phi(t)\sin qt dt\,.
\label{eq:A}
\end{equation}
The function $A(q)$ has the same properties as the corresponding function defined in the odd dimensional case \cite{hansen}~: it has neither zeros nor poles on the real axis, since by definition $s(q)$ has neither zeros nor poles for all $q$'s, $A(q)=A(-q)$, and $A(q)\rightarrow 1$ as $q\rightarrow \infty$. Therefore, in order to determine $\phi(t)$, we can use the formulation of Baxter \cite{baxter}, which was developed to solve the PY equation in odd dimensions. We use the Wiener-Hopf method by defining  \cite{baxter,hansen}
\begin{equation}
A(q)=\tilde{Q}(q)\tilde{Q}(-q)\,,
\label{eq:A1}
\end{equation}
where $\tilde Q(q)$ is an analytic function for $\Im (q)>0$. Following the same steps as in \cite{baxter,hansen}, it may be shown that $\tilde{Q}(q)$ can be written as
\begin{equation}
\tilde{Q}(q)=1-\rho \int_0^1Q(t)e^{iqt}dt\,.
\label{eq:Q}
\end{equation}
Substituting (\ref{eq:A1},\ref{eq:Q}) into Eq.~(\ref{eq:A}) gives
\begin{equation}
\frac{1}{q}\int_0^1 \phi(s)\sin qs ds= \int_0^1Q(s)e^{iqs}ds+ \int_0^1Q(s)e^{-iqs}ds-\rho \int_0^1ds\int_0^1ds'Q(s)Q(s')e^{iq(s-s')}\,.
\label{eq:phi}
\end{equation}
Multiplying (\ref{eq:phi}) by $\exp(-iqt)$, with $t>0$, and integrating with respect to $q$ from $-\infty$ to $\infty$ gives
\begin{equation}
\int_t^1   \phi(s)ds= 2Q(t)-2\rho \int_t^1 Q(s)Q(s-t)ds\,.
\label{eq:phi1}
\end{equation}
By setting $t=1$ in \eqref{eq:phi1} we see that
\begin{equation}
Q(1)=0\,.
\label{eq:bcQ}
\end{equation}
Moreover, Eq.~(\ref{eq:phi1}) can be simplified further by differentiating with respect to $t$ to give
\begin{equation}
\phi(t)= -2Q'(t)+2\rho \int_t^1 Q'(s)Q(s-t)ds \qquad 0\leq t\leq 1\,.
\label{eq:solphi}
\end{equation}
Therefore once the function $Q(t)$ is determined, the functions $\phi(t)$ and $c(r)$ may be determined from Eq.~(\ref{eq:solphi}) and Eq.~(\ref{eq:solc}) respectively.

Now let us work on the function $h(r)$. Since $\tilde{h}(q)=\tilde{h}(-q)$, one can write without loss of generality
\begin{equation}
q\tilde{h}(q)=\int_0^\infty \psi(t)\sin q t dt\,,
\label{eq:psi}
\end{equation}
where $\psi(t)$ is a real function and $h(r)$ is given in terms of $\psi(t)$ by
\begin{equation}
h(r)=\int_r^\infty \frac{\psi(t)}{\sqrt{t^2-r^2}} \frac{dt}{2\pi}\,.
\label{eq:solh}
\end{equation}
Substituting this form for $h(r)$ in condition~(\ref{eq:bc1}) yields
\begin{equation}
 \int_r^1 \frac{\psi(t)}{\sqrt{t^2-r^2}} \frac{dt}{2\pi}  =-1-\int_1^\infty \frac{\psi(t)}{\sqrt{t^2-r^2}} \frac{dt}{2\pi}\qquad 0<r<1\,.
\label{eq:cond1}
\end{equation}
Eq.~(\ref{eq:cond1}) is an Abel integral equation which can be inverted. It is easily shown that the inversion of the equation
\begin{equation}
\int_0^{\zeta_0} \frac{\sigma(\zeta)}{\sqrt{\zeta_0-\zeta}}\,d\zeta=\tau(\zeta_0)\qquad\zeta_0>0\,,
\end{equation}
yields
\begin{equation}
\sigma(\zeta)=\frac{d}{d\zeta}\left[\int_0^{\zeta}
\frac{\tau(\zeta_0)}{\sqrt{\zeta-\zeta_0}}\,\frac{d\zeta_0}{\pi}\right]\,.
\end{equation}
Therefore, using the change of variables $\zeta=1-t^2$ and $\zeta_0=1-r^2$, the inversion of Eq.~(\ref{eq:cond1}) gives
\begin{equation}
\psi(t)= \frac{-4t}{\sqrt{1-t^2}}\left[1+\int_1^\infty \frac{\sqrt{s^2-1}}{s^2-t^2}\psi(s)\frac{ds}{2\pi} \right] \qquad 0<t<1\,.
\label{eq:cond11}
\end{equation}
Eq.~(\ref{eq:cond11}) is an integral equation fixing $\psi(t)$ for $0<t<1$ as function of $\psi(t)$ for $t>1$.

Now, let us determine the relationship between $Q(t)$ and $\psi(t)$. Putting together the results of Eqs.~(\ref{eq:A},\ref{eq:A1},\ref{eq:Q},\ref{eq:psi}) in Eq.~(\ref{eq:O&Z1}) gives
\begin{equation}
\left[1-\rho \int_0^1Q(s)e^{iqs}ds\right]\left[1+\frac{\rho}{q} \int_0^\infty \psi(s)\sin qs ds\right]=\frac{1}{\tilde{Q}(-q)}\,.
\label{eq:psi1}
\end{equation}
Multiplying Eq.~(\ref{eq:psi1}) by $\exp(-iqt)$ with $t>0$ and integrating with respect to $q$ from $-\infty$ to $\infty$ gives~\cite{baxter,hansen}
\begin{equation}
4 Q(t)=
\int_0^\infty ds \psi(s) (\epsilon(s+t)+\epsilon(s-t)) -\rho \int_0^1ds Q(s) \int_0^\infty ds' \psi(s') (\epsilon(s'+t-s)+\epsilon(s'-t+s))\,,
\label{eq:psi2}
\end{equation}
where $\epsilon(x)=\mathrm{sign}(x)$. Using $\epsilon(x)=-1+2\Theta(x)$, with $\Theta(x)$ the Heaviside function, and differentiating Eq.~(\ref{eq:psi2}) with respect to $t$ gives
\begin{equation}
2Q'(t)+\psi(t) =\rho \int_0^1 Q(s) \bigl[\psi(t-s)-\psi(s-t)\bigr]ds\qquad t>0\,.
\label{eq:psi3}
\end{equation}
Recalling that $\psi(t)$ is defined for $t>0$ and that $Q(t)$ is defined for $0<t<1$, Eq.~(\ref{eq:psi3}) can be split into two integral equations as follows
\begin{eqnarray}
2Q'(t)+\psi(t)&=& \rho \int_0^t Q(s)\psi(t-s)ds-\rho \int_t^1 Q(s) \psi(s-t)ds\qquad \mbox{for}\qquad 0<t<1\,,
\label{eq:cond2}\\
\psi(t)&=& \rho \int_0^1 Q(s)\psi(t-s)ds\qquad \mbox{for}\qquad t>1\,.
\label{eq:cond3}
\end{eqnarray}

Our approach has reduced the PY problem for hard discs to the solution of the integral equations (\ref{eq:cond11},\ref{eq:cond2},\ref{eq:cond3}) with the additional boundary condition~(\ref{eq:bcQ}). We note that unlike the odd dimensional case \cite{leuth1}, here one cannot separate the problem of finding the direct correlation function $c(r)$ from that of finding the pair correlation function $g(r)$. This is because the behavior of $\psi(t)$ for $0<t<1$ is related to the behavior of $\psi(t)$ for $t>1$ through Eq.~(\ref{eq:cond11}). Although we were unable to find an analytical solution valid for all $\rho$, the numerical solution of these equations can be easily implemented. Before dealing with the numerical analysis, let us first consider the equation of state in the present formulation of the problem.

There are two methods used to calculate the equation of state when the radial distribution function, $g(r)$, is known. Without the assumptions made in deriving the PY equation, these two methods would yield the same equation of state. The difference in the equations of state calculated using these two methods therefore provides an estimation of the error made by using the PY approximation. The first equation of state is derived from the virial theorem and is given by \cite{hansen}
\begin{equation}
 \beta P^{(v)}=\rho+\frac{\pi}{2}\,\rho^2 \,g(1^+) \,.
 \label{eq:Pv}
\end{equation}
Using $ g(1^+) = - c(1^-)$ \cite{hansen} and Eqs.~(\ref{eq:asympt},\ref{eq:solphi}), the ``virial" equation of state becomes
\begin{equation}
 \beta P^{(v)}=\rho+\frac{\pi}{4}\,\rho^2 \,\lim_{t\rightarrow1^-} \sqrt{1-t^2}Q'(t) \,.
 \label{eq:Pv1}
\end{equation}
The second method uses the isothermal compressibility $\kappa_T$ which is given by~\cite{hansen}
\begin{equation}
\rho \beta^{-1}\kappa_T=\beta^{-1}\left(\frac{\partial \rho}{\partial P^{(c)}}\right)_T=s(q=0)\,.
\label{eq:comp}
\end{equation}
Using Eq.~\eqref{eq:A} and replacing $\phi(t)$ with $Q(t)$ by using Eq.~\eqref{eq:solphi}, one has
\begin{equation}
\frac{1}{s(0)}=1+2\rho\int_0^1\left[\rho\int_t^1Q(s)Q(s-t)ds-Q(t)\right]dt \, .
\label{eq:s(0)b}
\end{equation}
Integrating \eqref{eq:comp} and substituting for $s(0)^{-1}$ from \eqref{eq:s(0)b} we find the ``compressibility" equation of state
\begin{equation}
 \beta P^{(c)}=\rho+2\int_0^\rho 2\rho'\left\{\int_0^1\left[\rho'\int_t^1Q(s)Q(s-t)ds-Q(t)\right]dt\right\}d\rho'\,.
 \label{betapceq}
\end{equation}

\section{Numerical Procedure}

We found it simplest to implement an iterative procedure for solving the set of integral equations (\ref{eq:cond11},\ref{eq:cond2},\ref{eq:cond3}). First, note that for $\rho=0$, the exact solution is given by
\begin{eqnarray}
\psi(t)&=& \frac{-4t}{\sqrt{1-t^2}}\Theta(1-t) \,,\\
Q(t)&=&-2\sqrt{1-t^2}\Theta(1-t) \,,
\end{eqnarray}
and that for any $\rho$, one can write the functions $Q(t)$ and $\psi(t)$ as power series in $\rho$
\begin{eqnarray}
\psi(t) &=&\frac{-4t\Theta(1-t)}{\sqrt{1-t^2}}\sum_{i=0}^\infty \rho^i\Psi^{(i)}(t)+\sum_{i=1}^\infty \rho^i  \psi^{(i)}(t)\Theta(t-1)\Theta(i+1-t)\,.
\label{eq:psi(t)}\\
Q(t) &=&-2\sqrt{1-t^2}\sum_{i=0}^\infty \rho^i q^{(i)}(t)\,,
\label{eq:q(t)}
\end{eqnarray}
with the definition
\begin{eqnarray}
\Psi^{(0)}(t)&= &q^{(0)}(t)=1\,,\\
\psi^{(0)}(t)&=&0\,.
\label{zeroorder}
\end{eqnarray}
Using this power series representation, Eq.~(\ref{eq:cond11}) yields
\begin{equation}
\Psi^{(i)}(t)=\int_1^{i+1} \frac{\sqrt{s^2-1}}{s^2-t^2}\psi^{(i)}(s)\frac{ds}{2\pi}\qquad i\geq 1\,,
\label{eq:num31b}
\end{equation}
and Eq.~(\ref{eq:cond2}) becomes
\begin{equation}
(1-t^2)\,q'^{(i+1)}(t) -t\,q^{(i+1)}(t)+t\,\Psi^{(i+1)}(t)=- 2\sqrt{1-t^2} \int_0^1  \frac{(t-s)\sqrt{1-s^2}}{\sqrt{1-(t-s)^2}}\sum_{k=0}^{i} q^{(k)}(s)\Psi^{(i-k)}(|t-s|)ds\qquad i\geq 0\,.
\label{eq:num21c}
\end{equation}
The boundary condition $q^{(i)}(1)=\Psi^{(i)}(1)$ for $\forall i \geq 0$ is necessary to avoid a singularity in $q^{(i)}(t)$ as $t\rightarrow1$. Finally, Eq.~(\ref{eq:cond3}) gives
\begin{eqnarray}
\psi^{(i+1)}(t)&=& \sum_{k=0}^{i}\left[8 \int_{t-1}^1 \frac{s\sqrt{1-(t-s)^2}}{\sqrt{1-s^2}} q^{(k)}(t-s)\Psi^{(i-k)}(s)ds-2 \int_1^t  \sqrt{1-(t-s)^2}q^{(k)}(t-s)\psi^{(i-k)}(s)ds\right],\;1\leq t\leq2
\label{eq:num31c}\\
\psi^{(i+1)}(t)&=&-2 \sum_{k=0}^{i}\int_{t-1}^t \sqrt{1-(t-s)^2}q^{(k)}(t-s)\psi^{(i-k)}(s)ds,\; t> 2\,,
\label{eq:num31cc}
\end{eqnarray}

In devising a successful numerical scheme for this problem, it is important to note that several of the integrands have mild singularities, for example at $s=1$ in \eqref{eq:num31c} and at $s=1,t=1$ in \eqref{eq:num31b}. The singularity in \eqref{eq:num31c} may be dealt with simply by subtraction. To deal with the singularity in \eqref{eq:num31b}, we note that we may integrate \eqref{eq:num31b} once by parts and write
\begin{equation}
\Psi^{(i)}(t)=\sqrt{1-t^2}\int_1^{i+1}\arctan\sqrt{\frac{s^2-1}{1-t^2}}\frac{d}{ds}\left(\frac{\psi^{(i)}(s)}{s}\right)\frac{ds}{2\pi}-\int_1^{i+1}\sqrt{s^2-1}\frac{d}{ds}\left(\frac{\psi^{(i)}(s)}{s}\right)\frac{ds}{2\pi}.
\end{equation} This obviates the need to subtract the singularity and also suggests that we may write
\begin{equation}
\Psi^{(i)}(t)=\alpha^{(i)}+\delta^{(i)}\sqrt{1-t^2}+g^{(i)}(t),
\label{tempgdefn}
\end{equation} with $g^{(i)}(1)=0$, and
\begin{eqnarray}
\alpha^{(i)}&=&-\int_1^{i+1}\sqrt{s^2-1}\frac{d}{ds}\left(\frac{\psi^{(i)}(s)}{s}\right)\frac{ds}{2\pi}=\Psi^{(i)}(1),
\label{alphadefn}\\
\delta^{(i)}&=&{\lim_{t\rightarrow1}} \int_1^{i+1}\arctan\sqrt{\frac{s^2-1}{1-t^2}}\frac{d}{ds}\left(\frac{\psi^{(i)}(s)}{s}\right)\frac{ds}{2\pi}=-\tfrac{1}{4}\psi^{(i)}(1).
\label{betadefn}
\end{eqnarray}

Examining the equation for $q^{(i+1)}(t)$, \eqref{eq:num21c}, we see that the form assumed for \eqref{tempgdefn} suggests that
\begin{equation}
q^{(i)}(t)=\alpha^{(i)}+\gamma^{(i)}\sqrt{1-t^2}+f^{(i)}(t).
\end{equation} Here, the function $f^{(i)}(t)$ satisfies $f^{(i)}(1)=0$ and
\begin{equation}
(1-t^2)f'^{(i+1)}(t)+t\bigl[g^{(i+1)}(t)-f^{(i+1)}(t)\bigr]=2\sqrt{1-t^2}\bigl[tA^{(i+1)}(1)-A^{(i+1)}(t)\bigr],
\label{newqeqn}
\end{equation} where
\begin{equation}
A^{(i+1)}(t)\equiv \sum_{k=0}^{i}\int_0^1\frac{(t-s)\sqrt{1-s^2}}{\sqrt{1-(t-s)^2}}q^{(k)}(s)\Psi^{(i-k)}(|t-s|)~ds
\end{equation}
The constant $\gamma^{(i)}$ may be determined by considering the limit $t\rightarrow1$, which gives
\begin{equation}
\gamma^{(i)}=\tfrac{1}{2}\delta^{(i)}+A^{(i)}(1).
\end{equation}

We solve the system of equations by discretizing in space using steps of size $\Delta$. Integrals are evaluated using the trapezoidal rule and derivatives are calculated using forward-differencing, both of which are first-order accurate in space. Iterations proceed from $i=0$ using the values $\Psi^{(0)}$ and $q^{(0)}$, given in \eqref{zeroorder}, to determine $\psi^{(1)}(t)$. Using $\psi^{(1)}(t)$ in the discretized versions of \eqref{alphadefn} and \eqref{betadefn} allows us to determine $\Psi^{(1)}(t)$. Substituting these into the discretized \eqref{newqeqn}, $q^{(1)}(t)$ may be determined. The whole process is then iterated $N$ times, corresponding to determining the first $N$ terms in the series for the variables $\Psi(t)$, $q(t)$ and $\psi(t)$. The results presented here typically have $N=20$.

\section{Numerical results}

Let us begin by presenting our results for the virial coefficients and the equation of state. The virial coefficients, $B_i$, are defined by
\begin{equation}
\beta P=\sum_{i=1}^\infty B_i\rho^i.
\label{virialdefn}
\end{equation}
It is well-known that $B_1=1$ and $B_2=\pi/2$ are exact and are recovered within the PY approximation. The higher order virial coefficients can be calculated from the solution of the PY equation using the expressions in \eqref{eq:Pv1} and \eqref{betapceq}. For example, using Eqs.~(\ref{eq:q(t)},\ref{eq:Pv1}) yields
\begin{equation}
 B_i^{(v)}=\frac{\pi}{2} q^{(i-2)}(1) \qquad i\geq2\,,
 \label{eq:Pv2}
\end{equation}
Thanks to the iterative procedure presented above, these coefficients as well as the $B_i^{(c)}$, which are derived from the isothermal compressibility \eqref{betapceq}, are directly given by the numerical resolution of the problem.

Our numerical solution of the PY equation and calculation of the virial coefficients are only first order accurate in space and so we expect to accumulate errors at $O(\Delta)$. To counter this, we calculated each of the virial coefficients for several values of $\Delta$ and then extrapolated linearly its value at $\Delta=0$. For each virial coefficient the correlation coefficient was calculated to be $<-0.999$ confirming that the expected linear dependence of $B_i$ on $\Delta$ is indeed observed. The computed values of the first twenty coefficients for the series corresponding to $\beta P^{v}$ and $\beta P^{c}$ are presented in Table~\ref{virialcoeffs}. The values of the first ten virial coefficients for hard discs were determined recently using a ``hit or miss" Monte-Carlo integration algorithm in \cite{clisby06} and are reproduced in Table~\ref{virialcoeffs} for comparison.

\begin{table}
\centering \setlength{\tabcolsep}{0.8 em}
\begin{tabular}{cccc}
$i$ & $B_i^{(v)}$ & $B_i^{(c)}$& $B_i^{(MC)}$\\
\hline
3 &$1.930$&$1.930$&$1.930$\\
4 &$1.941$&$2.084$& $2.063$\\
5 &$1.795$&$2.090$&$2.031$\\
6  &$1.594$& $1.999$&$1.902$\\
7  &$1.381$& $1.852$&$1.726$\\
8 & $1.177$ & $1.677$&$1.534$\\
9   &$0.992$& $1.492$& $1.342$\\
10 &$0.829$&$1.309$&$1.162$\\
11 &$0.688$&$1.136$& ---\\
12 &$0.567$&$0.977$& ---\\
13 &$0.466$&$0.833$& ---\\
14 &$0.381$&$0.707$& ---\\
15 &$0.311$&$0.596$& ---\\
16  &$0.253$& $0.500$& ---\\
17  &$0.205$& $0.418$& ---\\
18 & $0.16(6)$ & $0.348$& ---\\
19  &$0.13(4)$& $0.28(9)$& ---\\
20  &$0.10(7)$& $0.23(9)$& ---
 \end{tabular}
\caption{Numerical values of the first twenty virial coefficients. The  $B_i^{(MC)}$ are the results from Monte Carlo calculations presented in~\cite{clisby06}. $B_i^{(v)}$ and $B_i^{(c)}$ are the values found from the solution to the PY equation using Eqs.~\eqref{eq:Pv2} and \eqref{betapceq} respectively.}
\label{virialcoeffs}
\end{table}

\begin{figure}
\centering
\includegraphics[height=6cm]{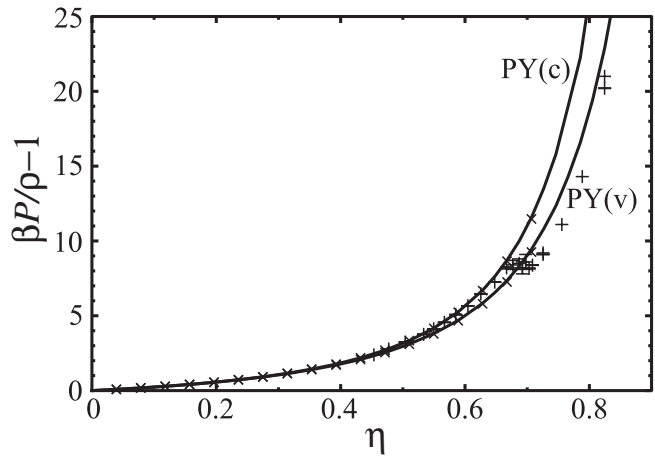}
\caption{Reduced pressure $\beta P/\rho-1$ as a function of the disk packing fraction, $\eta$. The results of molecular dynamics simulations reported previously \cite{hoover67,alder68} are shown as $+$ along with the results obtained from an earlier solution of the PY equation \cite{lado}, shown as $\times$. The two solid curves show the results of our calculations using the first $20$ virial coefficients, with the labels corresponding to the equation of state used to calculate them.}
\label{eqstate}
\end{figure}

As is clear from the definition \eqref{virialdefn}, the virial coefficients are important for determining the equation of state. The value of the pressure, $\beta P$, is therefore of considerable interest. Fig.~\ref{eqstate} shows the dependence of $\beta P$ on the packing fraction of the disks, $\eta$. This shows that there is good agreement between our results and previous ones using a direct approximate numerical resolution of PY equation \cite{lado}. Furthermore, this agreement is within the $2\%$ error estimate given in \cite{lado}.

\begin{figure}
\centering
\includegraphics[height=6cm]{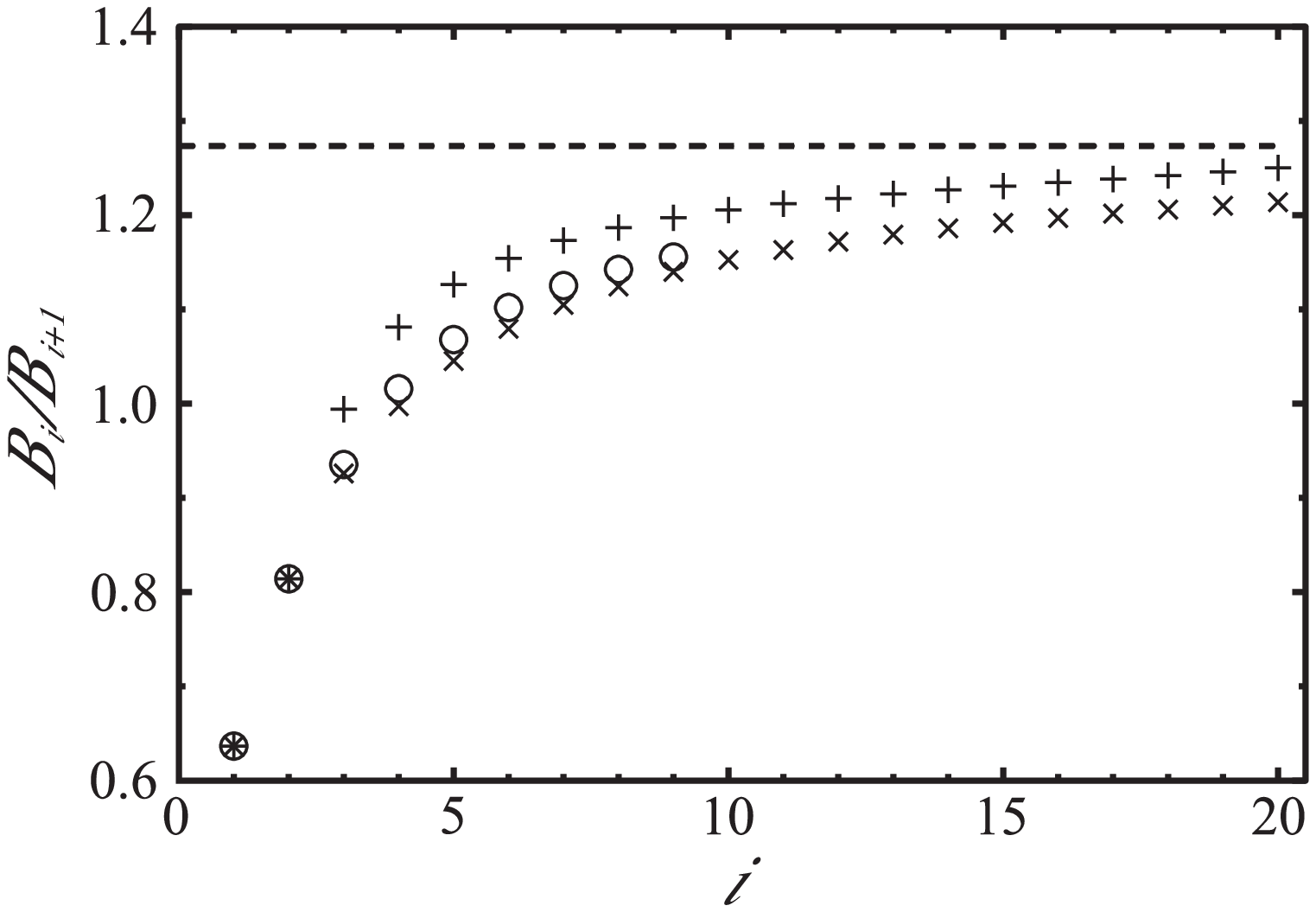}
\caption{The ratio between successive virial coefficients $B_i/B_{i+1}$ as a function of $i$. The ratios determined from the solution to the PY equation are denoted by $+$ ($\beta P^v$) and $\times$ ($\beta P^c$). These results sandwich the virial coefficients determined from Monte Carlo calculations given in \cite{clisby06}, which are represented by $\bigcirc$. The dashed line represents $B_i=4 B_{i+1}/\pi$, which appears to be an asymptote as $i\rightarrow\infty$.}
\label{clisbyfig}
\end{figure}

It is clear from Fig.~\ref{eqstate} that there is a divergence in $\beta P$ for relatively large $\eta$. The question of interest here is: at what density does this divergence occur? We are thus interested in determining the radius of convergence of the series for $\beta P$. To this end, we plot in Fig.~\ref{clisbyfig} the ratio between successive virial coefficients given in Table~\ref{virialcoeffs}. In particular, we observe that as $i\rightarrow\infty$
\begin{equation}
\frac{B_i}{B_{i+1}}\rightarrow \frac{4}{\pi}.
\end{equation}
Using d'Alembert's ratio test, this observation suggests that the series for $\beta P$ will converge absolutely provided that $\rho<4/\pi$ or, equivalently, $\eta<\eta_c=1$, the space filling density. This result shows that, similarly to the case of hard spheres ($d=3$), the PY equation for hard discs predicts no phase transition at intermediate density and thus fails at high densities.

\begin{figure}
\centering
\includegraphics[height=6cm]{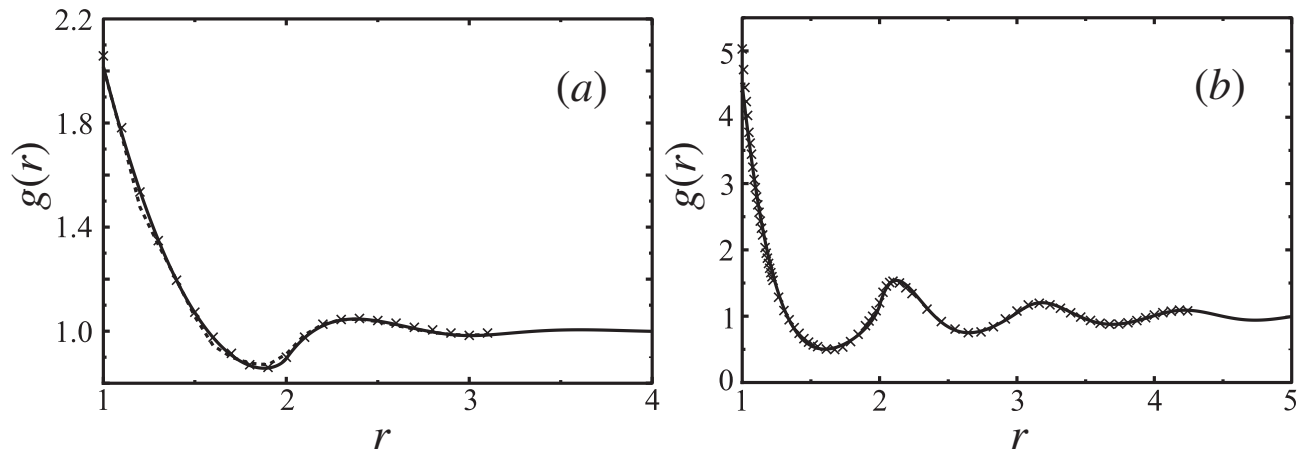}
\caption{The correlation function $g(r)$ computed from the PY equation (curves) and from different Monte Carlo simulations (crosses). (a) $\rho=0.462$: Our solution of the PY equation (solid curve) compared to the MC results of Chae \emph{et al.} \cite{numerics} (crosses) and their solution of the PY equation (dashed curve). (b) $\rho=0.794$: Our solution of the PY equation (solid curve) compared to the MC results of Wood \cite{wood} (crosses). Solid curves were calculated using the first $50$ terms of the series with $\Delta=0.00125$.}
\label{gr}
\end{figure}

Finally, the correlation function $g(r)$ may be calculated numerically from the values of $\psi^{(i)}$. Some typical results are shown in Fig.~\ref{gr} and compared with the available Monte Carlo results \cite{numerics,wood}. As can be seen, the overall agreement is very good, except in the vicinity of $r=1$. Also note that our solution of the PY equation in Fig.~\ref{gr}(a) agrees better with the Monte Carlo results than a previous numerical solution of the PY equation presented in \cite{numerics}.

\section{Discussion}
In this paper we developed a semi-analytic method to solve the PY equation for hard discs. At the heart of this approach is a reduction of the PY equation to a set of integral equations for two auxiliary functions $Q(s)$ and $\psi(s)$ as given by Eqs.~(\ref{eq:cond11},\ref{eq:cond2},\ref{eq:cond3}). The correlation functions and the equation of state can be determined easily from these auxiliary functions. We suggest an efficient iterative numerical method to solve these equations and determine the auxiliary functions. Using this method we are able to determine the values of the virial coefficients within the PY approximation. Furthermore, it allows for a comparison with the available $10$ virial terms of the full problem \cite{clisby06}. We also obtain results for the pair correlation function which compare well with the available MC calculations.

An advantage of this method is that it allows, in principle, calculations to arbitrary precision, and it could be interesting to obtain more virial coefficients by so doing. It could also be interesting to generalize this approach to polydisperse mixtures \cite{Leibowitz} and to higher even dimensions.

{\bf Acknowledgments}---This work was supported by EEC PatForm Marie Curie action (E.K.) and the Royal Commission for the Exhibition of 1851 (D.V.). Laboratoire de Physique Statistique is associated with Universities Paris VI and Paris VII.


\begin{thebibliography}{99}

\bibitem{hansen} J. P. Hansen and R. McDonald, {\it Theory of simple liquids} (Academic Press, New York, 2006).

\bibitem{PY} J. K. Percus and G. J. Yevick, Phys. Rev. {\bf 110}, 1 (1958).

\bibitem{baxter} R. J. Baxter, Aust. J. Phys. {\bf 21}, 563 (1968).

\bibitem{leuth1} E. Leutheusser, Physica A {\bf 127}, 667 (1984).

\bibitem{wertheim1} M. S. Wertheim, J. Math. Phys. {\bf 5}, 643 (1964).

\bibitem{wertheim2} M. S. Wertheim, Phys. Rev. Lett. {\bf 10}, 321 (1963).

\bibitem{thiele} E. Thiele, J. Chem. Phys. {\bf 39}, 474 (1963).

\bibitem{robles} M. Robles,  M. L. de Haro, and A. Santos A, J. Chem. Phys. {\bf 126}, 016101 (2007).

\bibitem{lado} F. Lado, J. Chem. Phys. {\bf 49}, 3092 (1968).

\bibitem{leuth2} E. Leutheusser J. Chem. Phys. {\bf 84}, 1050 (1986).

\bibitem{rosenfeld} Y. Rosenfeld, J. Chem. Phys. {\bf 87}, 4865 (1963).

\bibitem{hoover67} W. G. Hoover and B. J. Alder, J. Chem. Phys. {\bf 46}, 686 (1967).

\bibitem{numerics} D. G. Chae, F. H. Ree, and T. Ree, J. Chem. Phys. {\bf 50}, 1581 (1969).

\bibitem{alder68}  B. J. Alder, W. G. Hoover, and D. A. Young, J. Chem. Phys. {\bf 49}, 3688 (1968).

\bibitem{wood} W. W. Wood, J. Chem. Phys. {\bf 52}, 729 (1970).

\bibitem{numerics2} M. Bishop, P. A. Whitlock, and D. Klein, J. Chem. Phys. {\bf 122}, 074508 (2005).

\bibitem{clisby06} N. Clisby and B. M. McCoy, J. Stat. Phys. {\bf 122}, 15 (2006).

\bibitem{sned} I. N. Sneddon {\it The Use of Integral Transforms} (McGraw Hill, 1972).


\bibitem{OZ} L. S. Ornstein and F. Zernicke, Proc. Acad. Sci. Amsterdam {\bf 17}, 793 (1914).

\bibitem{grads} I. S. Gradshteyn and  I. M. Rhyzik {\it Table of Integrals, Series, and Products} (Academic Press, New York, 1994).

\bibitem{Leibowitz} J. L. Leibowitz, Phys. Rev. {\bf 133}, A895 (1964).


\end{thebibliography}
\end{document}